\documentclass[twocolumn,showpacs,preprintnumbers,amsmath,amssymb,floatfix]{revtex4}

\usepackage{graphicx} 
\usepackage{dcolumn}  
\usepackage{bm}       

\newcommand{\E}[1]{\langle#1\rangle}
\providecommand{\abs}[1]{\lvert#1\rvert}
\providecommand{\saddle}{\mathcal{S}} 
\providecommand{\mfmin}{{\min}}
\providecommand{\negev}{\mathcal{N}} 
\providecommand{\mt}{m_t}
\providecommand{\mtsq} {m_t^2}
\providecommand{\bdry}{{\mathfrak B}}
\providecommand{\thalf} {1/2}
\providecommand{\nuc}{ \mathcal{C}} 
\providecommand{\myeq}[1]{Eq.~\eqref{#1}}
\providecommand{\myfig}[1]{Fig.~\ref{#1}}
\renewcommand{\sp}{{\rm sp}}

\begin{document}

\title{Transient Nucleation near the Mean-Field Spinodal}
\author{A. O. Schweiger}
\email{aschweig@physics.bu.edu}
\author{K. Barros}
\author{W. Klein}
\affiliation{Center for Computational Science
             and Department of Physics, Boston University,
             Boston, MA, 02215}

\begin{abstract}
    Nucleation is considered near the pseudospinodal in a one-dimensional $\phi^4$
    model with a non-conserved order parameter and long-range
    interactions. For a sufficiently large system
    or a system with slow relaxation to metastable equilibrium, there is a
    non-negligible probability of nucleation occurring
    before reaching metastable equilibrium.  This
    process is referred to as transient nucleation.
    The critical droplet is defined to be the configuration
    of maximum likelihood that is dynamically balanced between
    the metastable and stable wells.
    Time-dependent droplet profiles and nucleation rates are derived,
    and theoretical results are compared to computer simulation.
    The analysis reveals a distribution of nucleation times with
    a distinct peak characteristic of a nonstationary nucleation
    rate.  Under the quench conditions employed, transient critical droplets
    are more compact than the droplets found in metastable equilibrium
    simulations and theoretical predictions.
\end{abstract}

\maketitle \date{\today}
    \section{Introduction}
    Standard nucleation theory predicts a single metastable equilibrium
    nucleation rate, which gives rise to an exponential distribution
    of nucleation times.
%
%
    A key assumption in the application of the
    standard theory is that the system is in metastable
    equilibrium.  In practice, any system that has undergone
    a quench relaxes into metastable equilibrium.
    During this relaxation, the system has a finite
    probability of initiating a decay to the stable phase
    via an isolated droplet.
    This process, referred to as transient nucleation,
    is of both practical and theoretical interest.
    Experimentally, transient nucleation has been observed in the
    laser melting of thin silicon films~\cite{Stiffler}, crystallization
    in amorphous alloys~\cite{Kumomi, Zhuang}, and liquid crystals~\cite{Sergan}.
    Theoretical work has concentrated on mean-field kinetic
    descriptions~\cite{Zeldovich, Kelton, Shneidman, Wu, Wyslouzil, Bartell, Maksimov, Kashchiev}.
    Studies of transient nucleation in $\phi^4$
    models~\cite{Boyanovsky, Seunarine, Gleiser, Intergroup}
    have shown a distribution of nucleation times with a distinct peak
    and exponential tail.

    In a mean-field system, the spinodal defines the limit of metastability.
    The spinodal divides the phase-diagram into two regions,
    one region exhibits a metastable phase while the other does not.
    In a system with finite but long-range interactions,
    the analogous limit of metastability is defined by the
    pseudospinodal. As the interaction range increases,
    statistics at the pseudospinodal converge to the corresponding
    mean-field spinodal values.  As the pseudospinodal is approached,
    the metastable system exhibits diverging susceptibility, correlation length,
    and correlation times that are characteristic of a critical
    point~\cite{Gulbahce,Novotny}.
    Nucleation near a pseudospinodal is structurally
    different from classical nucleation near the coexistence curve.
    In classical nucleation, droplets are
    compact fluctuations of the stable phase with a sharp
    interface separating the interior and exterior~\cite{Langer67}.
    However, near the pseudospinodal, droplets do not
    exhibit a well-defined interface nor a stable-phase interior~\cite{Unger84}.
    Evidence for nucleation near the pseudospinodal is observed in
    deeply quenched liquids~\cite{Yang,Trudu} and in
    solid-solid phase transitions~\cite{Cherne,Gagne}.
    Our objective is to study transient nucleation
    rates and transient critical-droplet profiles
    in a system with long-range interactions near
    the pseudospinodal.

    To make these ideas precise, we consider purely
    dissipative dynamics described by a Markovian Langevin
    equation that is a sum of a deterministic drift term
    and a stochastic noise term with zero mean.
%
%
    We define the metastable well to be
    the set of configurations that would follow the
    (noiseless) deterministic drift to a stationary configuration that 
    is not the global energy minimum.
    The metastable well boundary $\bdry$ consists of
    configurations that, upon random perturbation, would each drift to the
    metastable minimum with probability~\thalf.
    A similar definition is employed by Roy et al. and others~\cite{Roy,Monette,Iso1}.
    Transient critical droplets are characterized by the
    most probable configuration on the metastable well boundary
    at a given time $t$ since the quench.

    We define the {\it nucleation time} as
    the latest time after the quench such that the system configuration
    was located on the metastable well boundary; the
    {\it nucleating droplet} is the corresponding
    system configuration. A system has nucleated when
    there is a negligible probability of returning to the
    metastable well.

    Because the nucleation rate is an extensive quantity, transience
    is best observed when the relaxation to metastable
    equilibrium is slow and the system size is large.
%
    Our primary results are that in magnetic field quenches
    at a temperature below the critical temperature,
    the transient nucleation rate is adequately described by a
    quasi-equilibrium theory.
    Near the pseudospinodal, transient droplets are more compact
    than those that nucleate in metastable equilibrium.

    In Sec.\ II, we introduce the model and review the
    field theoretic treatment of nucleation near the mean-field spinodal,
    the basis for the subsequent analysis.  Section III considers the
    time-dependent likelihood of non-equilibrium states
    characterizing relaxation to metastable equilibrium.
    In Sec.\ IV, we find the time-dependent nucleation rate
    and transient critical droplet as a perturbation about the metastable
    equilibrium critical droplet.  In Sec.\ V, we compare our theoretical
    treatment to results from computer simulation.
    Section VI interprets and compares our results to
    previous work.
    \section{Metastability Near the Mean-Field Spinodal}
    We consider a one-dimensional
    ferromagnetic system with long-range interactions
    prepared in equilibrium with
    an initially negative external field,
    $h_{initial} < 0$.
    At time $t = 0$, the system is quenched
    and the external magnetic field $h$ is set to a positive value.
    The system then evolves toward metastable equilibrium.
    At some point after the quench, it will decay to the stable
    phase via nucleation.
%
%

    The potential of the system is given by
    \begin{equation}
    \label{FullPotentialEquation}
    V(\phi) = \epsilon \phi^2 + u \phi^4 - h \phi,
    \end{equation}
    where $\epsilon < 0$ and $u > 0$.
    For systems of interaction range $R$, the
    Ginzburg-Landau-Wilson Hamiltonian is
    \begin{equation}
    H[\phi] = \int dx \left [ \frac{R^2}{2} \left (\frac{d \phi}{dx}\right )^2 + V(\phi(x)) \right ].
    \end{equation}
    We assume purely dissipative dynamics, given by the time-dependent Ginzburg-Landau
    equation,
    \begin{equation}
    \label{Langevin}
    \frac{\partial \phi}{\partial t}  = -\frac{\delta H}{\delta \phi} + \eta
        = R^2 \frac{\partial^2 \phi}{\partial x^2}  - 2 \epsilon \phi - 4 u \phi^3 + h + \eta,
    \end{equation}
    where $\eta = \eta(x,t)$ is a zero-mean white noise with
    $\langle \eta(x,t) \eta(x',t') \rangle = 2 \beta^{-1} \delta(x-x') \delta(t-t')$, and $\beta$ is
    the inverse temperature.
    The mean-field metastable well vanishes when the applied
    magnetic field is the spinodal field, defined by
    $h_{\sp}^2 = 8 \abs{\epsilon}^3/\,27u$.
    The corresponding mean-field spinodal magnetization
    satisfies $\phi_{\sp}^2 = \abs{\epsilon} / \, 6u$.
    These quantities are obtained from setting ${dV/d\phi = d^2 V/d\phi^2 =0}$.

    Under the given quench conditions, the spinodal field
    is positive,
    ${h_{\sp} \gtrsim h > 0}$, and the mean-field spinodal magnetization
    is negative, ${\phi_{\sp} < 0}$.
    For convenience, we introduce ${\Delta_h = h_{\sp} - h}$, a measure of the depth
    of the metastable well.

    We expand the potential about the spinodal magnetization,
    $\phi_{\sp}$, retaining terms up to third order~\cite{Unger84, Buttiker},
    \begin{equation}
    \begin{split}
    V =& \; \phi_{\sp} \left(\Delta_h - \frac{3}{8} h_{\sp}\right) + \Delta_h (\phi - \phi_{\sp} ) \\
       & - \frac{1}{2} \frac{h_{\sp}}{\phi_{\sp}^2} (\phi - \phi_{\sp} )^3 + O((\phi - \phi_{\sp} )^4).
    \end{split}
    \end{equation}
    We concentrate on nucleation in the neighborhood of the metastable well.
    Near the spinodal, the details of the potential for large positive
    magnetizations has a negligible effect on nucleation.
    Therefore we drop the quartic term in the subsequent analysis.
    Within this cubic approximation, the location of the potential minimum is
    \begin{equation}
    \label{MetastableMinimum}
    \phi_{\mfmin} = \phi_{\sp}\left[1 + \sqrt{\frac{2 \Delta_h}{3 h_{\sp}}}\;\right] < 0.
    \end{equation}
    We introduce the shifted field, $\psi(x) = \phi(x) - \phi_{\mfmin}$, so that
    the minimum of the mean-field potential occurs precisely at $\psi = 0$.
    We rewrite the truncated potential in terms of the shifted field
    and drop an overall constant to obtain
    \begin{equation}
    \label{CubicPotential}
    V(\psi) = - \frac{\sqrt{6 h_{\sp} \Delta_h}}{2 \phi_{\sp}} \, \psi^2 - \frac{1}{2} \frac{h_{\sp}}{\phi_{\sp}^2} \psi^3
            = a \psi^2 - b \psi^3,
    \end{equation}
    where the variables
    ${a = -\sqrt{6 h_{\sp} \Delta_h} / \, 2 \phi_{\sp} > 0}$
    and ${b = a^2 / \, 3 \Delta_h > 0}$ are introduced to simplify the notation.
    Near the mean-field spinodal, the evolution of the field in
    the neighborhood of the metastable well is given by a Langevin equation,
    \begin{equation}
    \label{CubicLangevin}
    \frac{\partial \psi}{\partial t} = R^2 \frac{\partial ^2 \psi}{\partial x^2} - 2 a \psi + 3 b \psi^2 + \eta.
    \end{equation}
    We define the stationary configurations by
    setting the time derivative and the noise term to zero,
    \begin{equation}
    \label{FixedPointEqn}
    0 = R^2 \frac{d^2 \psi}{dx^2} - 2 a \psi + 3 b \psi^2.
    \end{equation}
%
%
    Equation~\eqref{FixedPointEqn} has two spatially uniform solutions,
    the stable configuration, $\psi_{\mfmin}(x) = 0$, and an unstable
    configuration.
    There exists another non-uniform stationary configuration $\psi_{\saddle}(x)$
    that corresponds to the profile of the critical droplet in metastable
    equilibrium~\cite{Unger84,Langer67}.
    Because the droplet can appear anywhere in the system,
    we fix the center of the droplet at the origin.
    A droplet must be symmetric, $\psi_{\saddle}(x) = \psi_{\saddle}(-x)$,
    and cannot exhibit a sharp peak at its center, $\psi^{\prime}_{\saddle}(0) = 0$.
    At large distances, the droplet profile approaches the
    metastable background, $\lim_{x \rightarrow \pm \infty} \psi_{\saddle}(x) = 0$.
    Solving \myeq{FixedPointEqn}, the non-uniform solution is
    \begin{equation}
    \label{FixedPoint}
    \psi_{\saddle} = \frac{a}{b} \cosh^{-2} \left( x \, \sqrt{\frac{a}{2 R^2}}  \; \right).
    \end{equation}
%
%
    This stationary configuration is unstable:
    assuming noiseless dynamics, a random perturbation
    about $\psi_{\saddle}$ would cause the system to return to its
    metastable well with probability~\thalf.
    Linear stability analysis about the non-uniform solution
    reveals a single negative eigenvalue, $-5a/2$. The corresponding
    eigenvector gives the initial direction of growth (or decay)
    of the metastable equilibrium critical
    droplet~\cite{Buttiker,Langer67,Unger85},
    \begin{equation}
    \label{Eigenvector}
    \psi_{\negev} = \cosh^{-3} \left( x \, \sqrt{\frac{a}{2 R^2}} \; \right).
    \end{equation}
    The amplitude of the growth eigenvector is concentrated at its center;
    a spinodal critical droplet grows by first filling its center~\cite{Unger85}.
    \section{Characterizing Relaxation to Metastable Equilibrium}
    In this section, we estimate the time-dependent
    likelihood of field configurations, which
    is required to find transient critical-droplet
    profiles~\cite{Langer69, Binder73, Reguera}.
%
    After the quench, the time evolution of the probability
    is given by the functional Fokker-Planck equation.
    We sidestep the formal solution and use a quasi-static mean-field theory.
    We assume at a time $t$ since
    the quench, the likelihood of a configuration in the
    metastable well can be approximated by
    \begin{equation}
    \label{Probability}
    P[\psi | t]
        \approx C \exp \left \lbrace -\beta \int dx \left [ \frac{R^2}{2} \left (\frac{d\psi}{dx} \right)^2 + F_t \right ] \right \rbrace,
    \end{equation}
    where $C$ is an undetermined constant, and ${F_t = F_t(\psi)}$
    is a time-dependent polynomial in the field.
    At $t = 0$, \myeq{Probability} should agree with
    the equilibrium distribution prior to the quench.
    At long times, we must recover the metastable potential,
    \begin{equation}
    \label{LongTimeLimit}
    \lim_{t \to \infty} F_t = V(\psi) = a \psi^2 - b \psi^3.
    \end{equation}
    Because the quench only changes the external magnetic field,
    we assume that $F_t$ differs from the metastable equilibrium potential given in
    \myeq{CubicPotential}
    by a time-dependent effective external field, $J(t)$,
    \begin{equation}
    \label{TransientPotential}
    F_t = a \psi^2 - b \psi^3 + [h - J(t)] \psi.
    \end{equation}
    In the long time limit, \myeq{LongTimeLimit} implies
    that $J(t) \rightarrow h$.
    We expect that the quartic term can still be neglected since this
    approximation will be used when $t$ is close to $t_{eq}$, when the system is in
    metastable equilibrium.

    We choose $J(t)$ so that the density given in \myeq{Probability}
    is consistent with the time evolution of the mean-field magnetization.
    In the mean-field limit, all fluctuations are suppressed
    and the system is described by a single scalar order parameter,
    $\psi(x,t) \rightarrow m_t$.
    Within the spinodal approximation, the time evolution of the
    mean-field system is governed by the ordinary differential equation,
    \begin{equation}
    \frac{d}{dt} \mt = -2 a \mt + 3 b \mtsq.
    \end{equation}
    The time evolution of the mean-field magnetization is,
    \begin{equation}
    \label{TransientMetastableMinimum}
        \mt = \frac{2 a m_0}{3 b m_0 + (2 a  - 3 b m_0)\exp(2at) },
    \end{equation}
    where the initial condition $m_0$ denotes the prequench
    shifted mean-field magnetization.
    In order that \myeq{Probability} reproduce the mean-field
    dynamics, $F_t$ evaluated at $m_t$ must be a minimum,
    \begin{equation}
    \frac{d F_t}{d \psi} \bigg \vert_{\psi = m_t} = 0.
    \end{equation}
    The expression for the time-dependent effective field
    is
    \begin{equation}
    \label{EffectiveField}
    J(t) = 2 a \mt - 3 b \mtsq + h.
    \end{equation}
    Formally, Eqs.~\eqref{Probability},~\eqref{TransientPotential}, and~\eqref{EffectiveField} define
    the first term of a large-$R$ asymptotic expansion of the solution to the
    functional Fokker-Planck equation in the metastable well near the mean-field spinodal.

    Both the location of the minimum of $F_t$ and its curvature
    change as a function of $t$. For convenience, we
    introduce ${\Delta_J = h_{\sp} - J(t)}$ and
    \begin{equation}
    \label{CurvatureEvolution}
    A_t = \frac{1}{2} F_t^{\prime \prime}(m_t) = a\sqrt{\Delta_J / \Delta_h},
    \end{equation}
    a measure of the curvature of $F_t$ about its minimum.
    At long times, the mean-field magnetization approaches
    its metastable equilibrium value, $\mt \rightarrow 0$,
    which implies that $A_t \rightarrow a$ in the same limit.

    \begin{figure}
    \includegraphics[height=2.5in,width=3in]{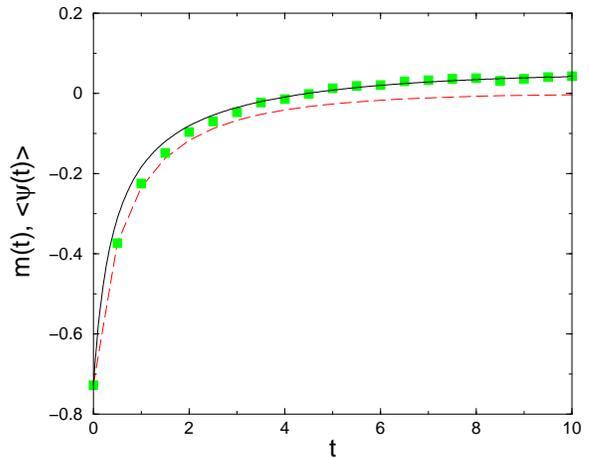}
    \caption{\label{Sample}
            The mean transient magnetization prior to nucleation (solid line) obtained
            from the solution of \myeq{Langevin} compared to the average
            obtained from simulating \myeq{QuasiLangevin} in metastable equilibrium
            (squares).  Both curves rise above the metastable mean-field minimum
            (broken line) given by \myeq{TransientMetastableMinimum}.}
    \end{figure}

    For a fixed time $t$, we define a corresponding fictitious
    Langevin equation,
    \begin{equation}
    \label{QuasiLangevin}
    \frac{\partial \psi(x, s)}{\partial s} = R^2 \frac{\partial^2 \psi}{\partial x^2} - F_t^{\prime}(\psi) + \eta(x, s),
    \end{equation}
    where the coordinate $s$ is distinct from the
    time $t$ since the quench.  Once in metastable equilibrium,
    \myeq{QuasiLangevin} samples states from \myeq{Probability}.
    We note that the states sampled by \myeq{QuasiLangevin}
    must be restricted to those within
    the metastable well of the complete dynamics of \myeq{Langevin}.
    For example, configurations that would nucleate \myeq{QuasiLangevin}
    must be discarded because they are not in the metastable well of the underlying dynamics.
    With this restriction, we use \myeq{QuasiLangevin} to compare the
    statistics of \myeq{Probability} at a fixed time $t$ to those generated
    from simulation of the complete dynamics in \myeq{Langevin}.
    Figure~\ref{Sample} compares the ensemble-averaged transient
    magnetization obtained from \myeq{Langevin} to the metastable equilibrium
    magnetization sampled using \myeq{QuasiLangevin} for various $t$.

%
%
    \begin{figure*}
    \begin{minipage}[t]{3.2in}
    \includegraphics[height=2.5in,width=3in]{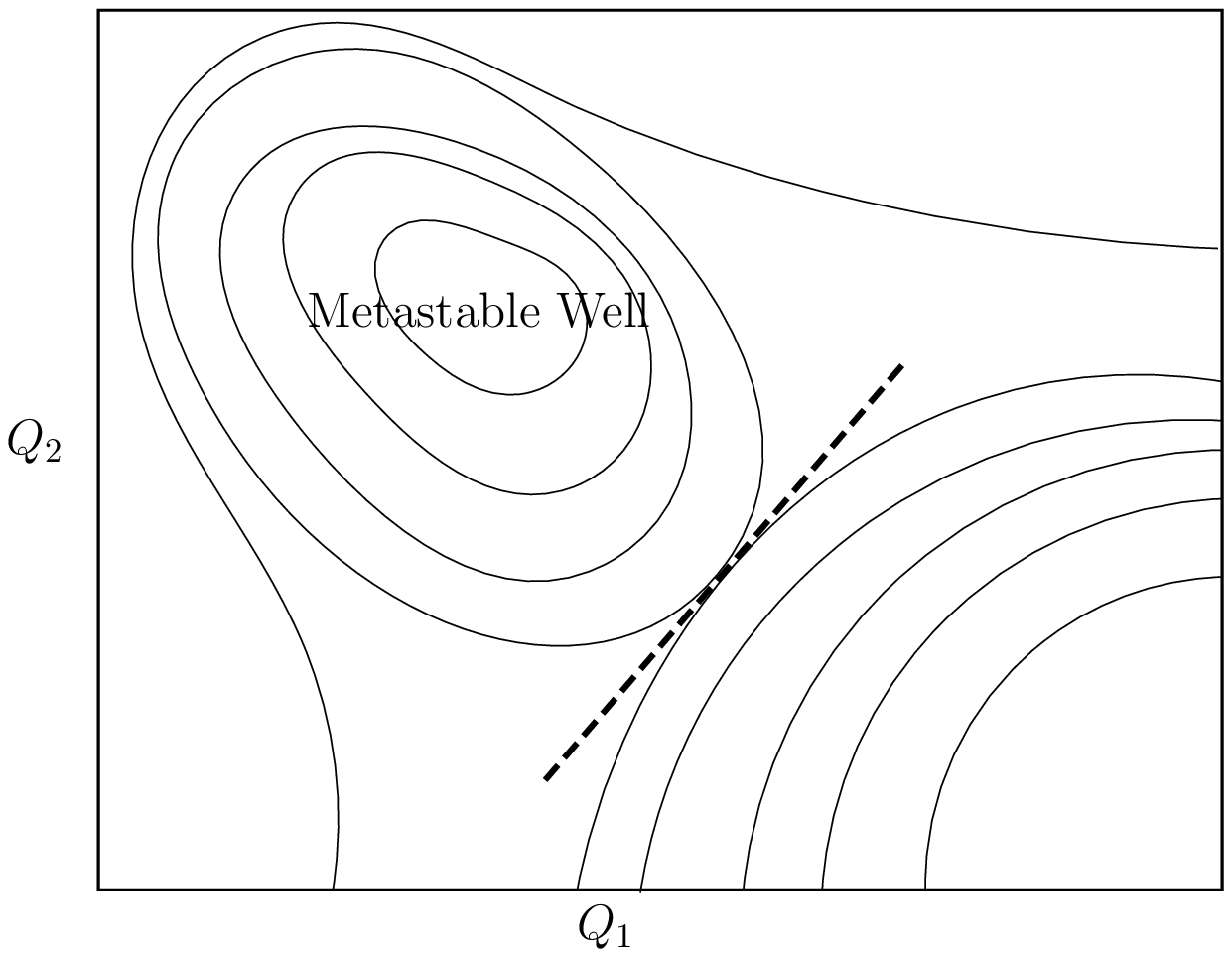}
    \caption{\label{Schematic}
            Schematic equipotential curves for a model
            system with two arbitrary dynamical variables, $Q_1$ and $Q_2$.
            The broken line illustrates the linearization of
            the metastable well boundary.}
    \end{minipage}
    \begin{minipage}[t]{0.1in}
    \hfill
    \end{minipage}
    \begin{minipage}[t]{3.2in}
    \includegraphics[height=2.5in,width=3in]{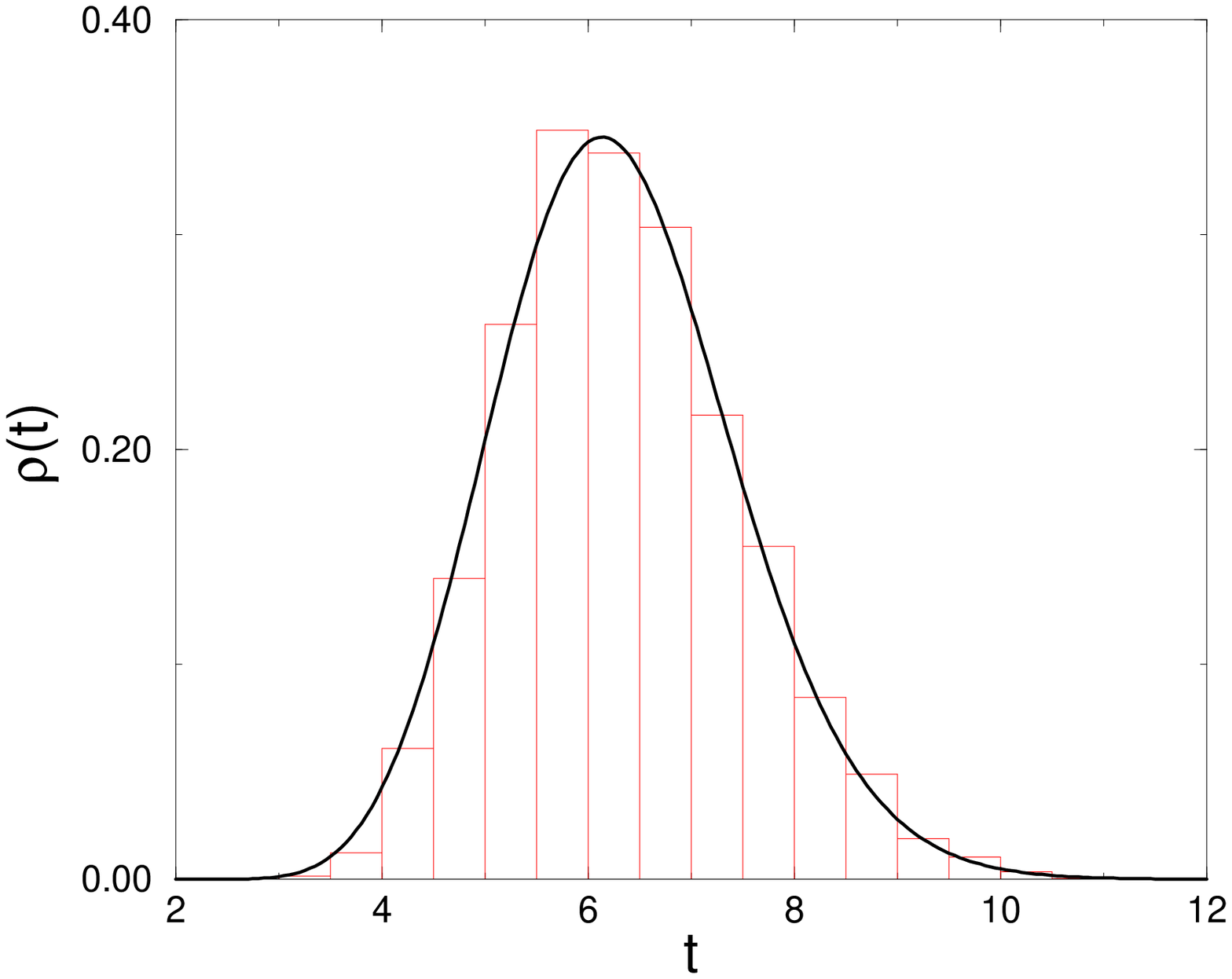}
    \caption{ \label{Times} The theoretical distribution of nucleation times computed
              from \myeq{NucleationDistribution} (solid line)
              compared to the normalized histogram of nucleation times obtained
              from direct simulation of \myeq{Langevin}.
              }
    \end{minipage}
    \end{figure*}

    To analytically estimate the transient magnetization of \myeq{Probability},
    we compute the stationary spatial average of \myeq{QuasiLangevin},
    \begin{equation}
    \label{MeanFieldMoments}
    0 = -2 a \E{\psi} + 3 b \E{\psi^2} + J(t) - h,
    \end{equation}
    which relates the first and second moments.
    The second moment can be approximated by the second moment of
    the Gaussian theory obtained by setting $b=0$,
    \begin{equation}
    \label{SecondMoment}
    \E{\psi^2} \approx \mtsq + \frac{1}{2 \beta R \; (2 A_t)^{1/2}}.
    \end{equation}
    Equations~\eqref{MeanFieldMoments} and~\eqref{SecondMoment} yield
    an estimate of the magnetization as a function of
    $A_t$, defined in \myeq{CurvatureEvolution},
    \begin{equation}
    \label{MeanMagnetization}
    \E{\psi} \approx \mt + G_t^{-1} > \mt,
    \end{equation}
    where,
    \begin{equation}
    \label{GinzburgParameter}
    G_t = 4 \beta R \Delta_h a^{-1} \sqrt{2 A_t} \geq G_\infty.
    \end{equation}
    This estimate of the magnetization
    also agrees with the average magnetization obtained
    from simulation, shown in \myfig{Sample}.
%
%
%
%

    \section{Transient Nucleation}
    \begin{figure*}
    \begin{minipage}[t]{3.2in}
    \includegraphics[height=2.5in,width=3in]{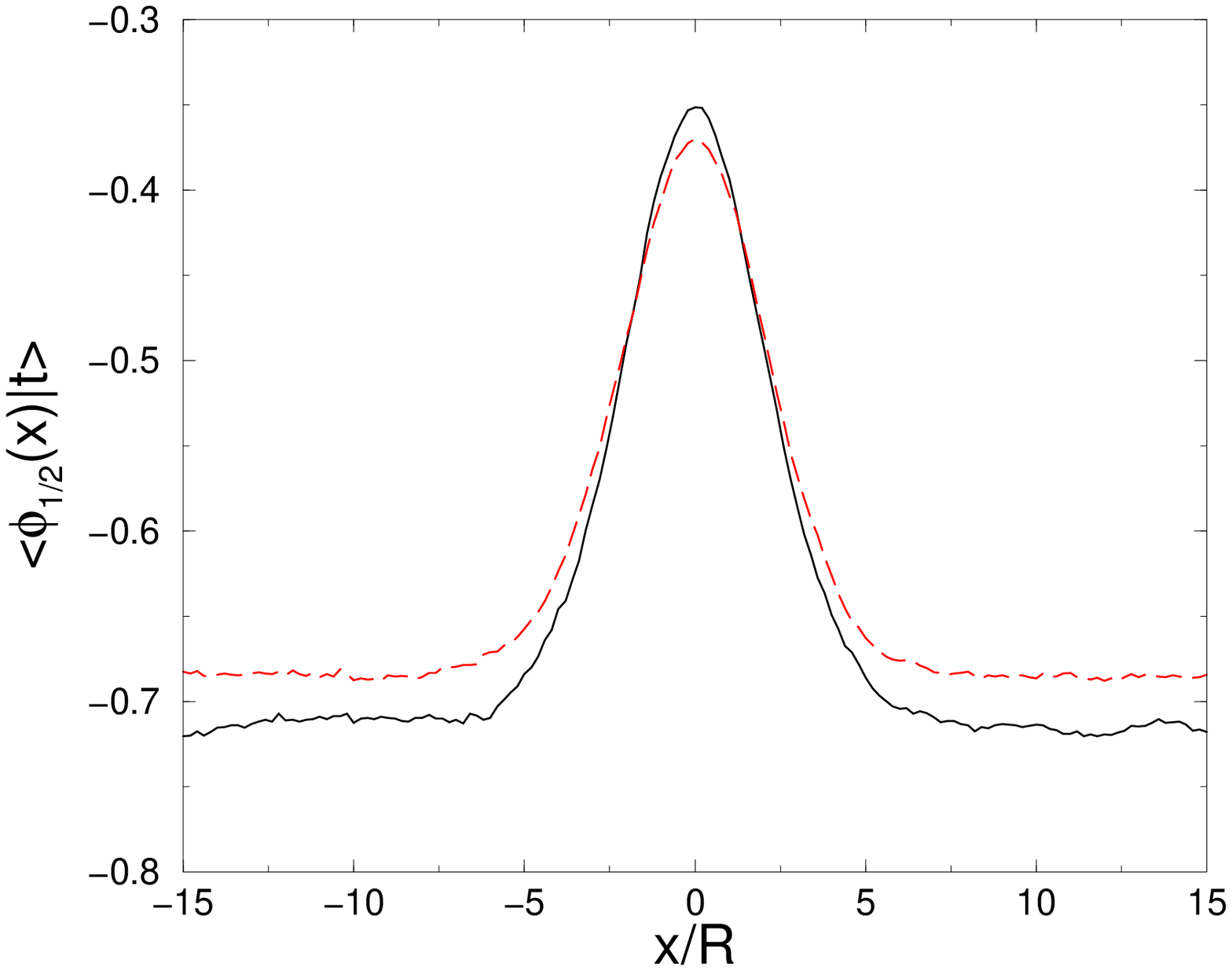}
    \caption{  \label{SimulationDroplets} Ensemble-averaged nucleating-droplet magnetization
         profiles from simulation at $t = 4.5$ (solid line) and $t = 7.5$ (broken line).
         }
    \end{minipage}
    \begin{minipage}[t]{0.1in}
    \hfill
    \end{minipage}
    \begin{minipage}[t]{3.2in}
    \includegraphics[height=2.5in,width=3in]{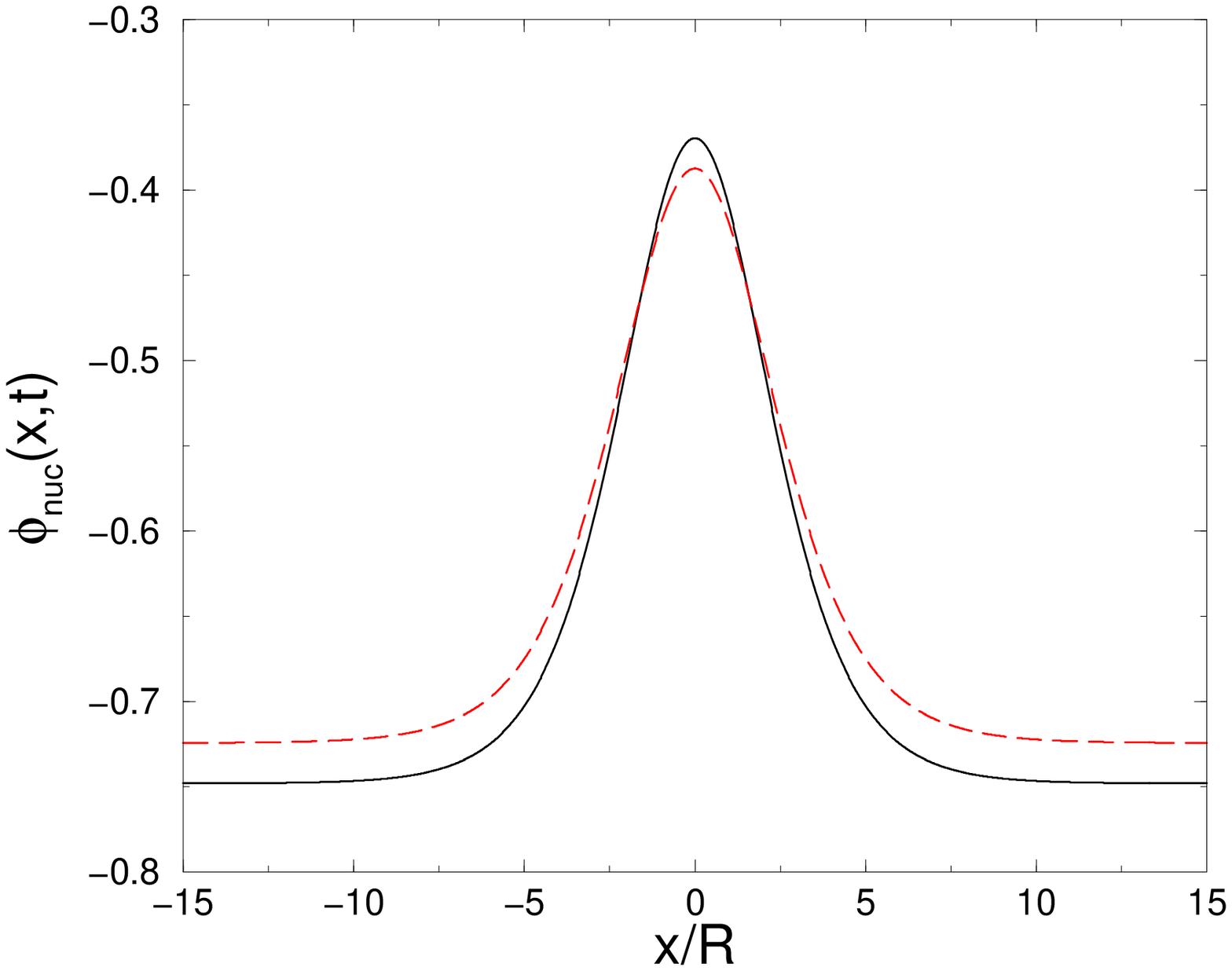}
    \caption{   \label{TheoryDroplets} Theoretical critical-droplet magnetization profiles from
        \myeq{Profile} computed at $t = 4.5$ (solid line) and $t = 7.5$ (broken line).  }
    \end{minipage}
    \hfill
    \end{figure*}
    At a time $t$ after the quench, the transient critical droplet $\psi_{\nuc}(x,t)$
    is determined by maximizing the probability functional in \myeq{Probability}
    constrained to the metastable well boundary $\bdry$,
    \begin{equation}
    \label{NucleatingDroplet}
    \psi_{\nuc}(x, t) = \max_{\psi \in \bdry } P[ \psi | t].
    \end{equation}
%
%

    Note that the dynamically
    defined stationary configuration $\psi_{\saddle}$ given in \myeq{FixedPoint}
    lies on the boundary.
    Furthermore, we assume that our transient
    critical droplets resemble the dynamical fixed point, $\psi_{\saddle}$.
    In this limit, we approximate the metastable well boundary
    by a hyperplane~\cite{Roy} that is normal to the growth eigenvector
    $\psi_{\negev}$, given in \myeq{Eigenvector}.  Figure~\ref{Schematic}
    gives a schematic of the approximation for a two-dimensional system.
    Within this approximation, a boundary configuration $\psi_{\bdry}$
    satisfies
%
%
    \begin{equation}
    \label{Constraint}
    \int dx \; (\psi_{\bdry} - \psi_{\saddle}) \psi_{\negev} = 0.
    \end{equation}
    To find the configuration of maximum likelihood that
    satisfies the constraint, we introduce the Lagrange multiplier
    $\lambda_t$ and extremize the functional,
    \begin{equation}
    \int dx \left [ \frac{R^2}{2} \left (\frac{d \psi}{dx} \right)^2 + F_t(\psi)
        - \lambda_t \psi \; \psi_{\negev} \right ].
    \end{equation}
%
%
    Centered at zero, the time-dependent critical-droplet profile solves the
    Euler-Lagrange equation,
%
%
    \begin{equation}
    R^2 \frac{d^2 \psi_{\nuc}}{dx^2} = 2 a \psi_{\nuc} - 3 b \psi_{\nuc} - J(t) + h - \lambda_t \psi_{\negev},
    \end{equation}
    where $\lambda_t$ is chosen so that $\psi_{\nuc}$ satisfies the constraint in \myeq{Constraint}.
    For large $t$, the transient solution reduces to the dynamical fixed point,
    ${\lim_{t \rightarrow \infty} \psi_{\nuc} = \psi_{\saddle}}$.
    At earlier times, the equation and constraint can be solved numerically.
    Because we have assumed the transient critical droplet resembles the
    fixed point, we expect $\lambda_t \ll 1$.  This
    suggests a solution as a perturbation in $\lambda_t$.
    First, we take $\lambda_t = 0$ and find the analog to \myeq{FixedPoint},
    \begin{equation}
    \psi_{\nuc}^0(x, t) = \mt + \frac{A_t}{b} \cosh^{-2} \left ( x \sqrt{\frac{A_t}{2 R^2}} \right ).
    \end{equation}
    Generally, the perturbation series is
    \begin{equation}
    \label{Profile}
    \psi_{\nuc}(x, t) = \psi_{\nuc}^0(x, t) - \lambda_t \; y(x, t) + O(\lambda_t^2).
    \end{equation}
    The resulting differential equation for $y(x)$ is
    \begin{equation}
    \begin{split}
    R^2 \frac{d^2 y}{dx^2} =  \, & 2 \; y(x) A_t \bigl[ 1 - 3 \cosh^{-2}(x\sqrt{A_t/2R^2}) \bigr ] \\
        & + \cosh^{-3}( x \sqrt{a/2 R^2} ).
    \end{split}
    \end{equation}
    We retain terms to first order in $\lambda_t$, and substitute \myeq{Profile}
    into the constraint to find the following equation for the multiplier $\lambda_t$,
    \begin{equation}
    \label{Multiplier}
    \lambda_t = \frac{ \int dx \left ( \psi_{\nuc}^0(x, t)  - \psi_{\saddle} \right ) \psi_{\negev}}
                  {  \int dx \; y(x) \; \psi_{\negev} }.
    \end{equation}
    The effect of the constraint is to decrease the amplitude
    of the transient critical droplet.
%
%
%
%
%
%

%
%
%
%
%
%

    With the transient droplet profiles characterized, we
    estimate the time-dependent nucleation rate.
    Given a time $t$ before any nucleation event,
    there is unit probability of being in the metastable well.  
    With this normalization, the nucleation rate is proportional
    to the probability to realize the transient critical droplet.
    We approximate this normalized probability by the
    relative likelihood of the transient critical droplet
    over the most likely metastable configuration,
    the uniform field with magnetization $m_t$,
    \begin{equation}
    \Gamma(t) \sim \frac{P[\psi_{\nuc}(x,t) \; | \; t \,]}
        {P[\psi = m_t \; | \; t \,]}
    \end{equation}
    This rate estimate is simplified by defining the shifted effective potential,
    \begin{equation}
    \begin{split}
    \Delta F_t(\psi) & = F_t(\psi) - F_t(m_t) \\
                     & = A_t (\psi - \mt)^2 - 3b(\psi - \mt)^3.
    \end{split}
    \end{equation}
    With this definition, the transient nucleation rate is,
    \begin{equation}
    \Gamma_t  = \Gamma_0 \exp \left \lbrace -\beta \int dx \left [ \frac{R^2}{2} \left (\frac{d\psi}{dx} \right)^2 + \Delta F_t \right ] \right \rbrace,
    \end{equation}
    where $\Gamma_0$ is a prefactor that depends on the details of the
    system~\cite{Prefactor}. For the quench considered here,
    $\Gamma(t)$ is a strictly increasing function of time. The
    experimentally accessible nucleation time distribution is given by
    \begin{equation}
    \label{NucleationDistribution}
    \rho(t) = \Gamma(t) \; \exp \left \lbrace -\int_0^t dt \; \Gamma(t) \right \rbrace.
    \end{equation}
\section{Numerical Calculations}
%
    We simulated \myeq{Langevin} for a periodic system of length
    $L = 10^4$ where ${\epsilon = -5/9}$, ${u = 1/4}$, ${R = 10}$,
    and  ${\beta = 10}$. Prior to the quench, the system is prepared
    in equilibrium with an external field $h_{initial} = -1.40$ and
    shifted magnetization ${m_0 = -0.727}$.
    After the quench, the applied field is set to ${h = 0.430}$.
    The distance to the mean-field spinodal is
    ${\Delta_h = 0.021}$, and the location of the metastable minimum
    is given by ${\phi_{\mfmin} = -0.715}$.
    The corresponding parameters in the cubic approximation
    are ${a = 0.195}$ and ${b = 0.609}$.

    Throughout the run, the configuration of the system
    is saved periodically. Once the system has nucleated,
    we search for the nucleating droplet and nucleation time.
    We load a saved configuration and find
    the latest configuration that, upon perturbation, drifts to
    the stable phase with probability~\thalf~\cite{Monette}.
    We do not consider
    runs where multiple droplets appear at the nucleation time.
    The nucleation times
    were binned and compared to the theoretical results
    (see \myfig{Times}) where the free parameter $\Gamma_0$ was
    chosen to produce the best fit.  Ensemble-averaged nucleating
    droplet profiles for two different times are plotted
    in \myfig{SimulationDroplets}. The corresponding theoretical
    critical droplets obtained from \myeq{Profile} are plotted in
    \myfig{TheoryDroplets}.

    Figures~\ref{SimulationDroplets} and~\ref{TheoryDroplets}
    differ slightly because they represent different statistical quantities.
    The cubic term in \myeq{Probability} give rise to a skew in
    the distribution.  Consequently, average and
    extremal quantities do not agree.  The same effect is demonstrated in
    \myfig{Sample} and \myeq{MeanMagnetization}, where the transient system
    magnetization is found to differ from the mean-field magnetization.
    For distributions that are sufficiently peaked, we expect
    the discrepancies to vanish~\cite{Unger84, Langer67}.
    This occurs when the metastable-equilibrium Ginzburg parameter $G_{\infty}$
    given in \myeq{GinzburgParameter} is large.
    In the case considered here, the Ginzburg parameter is finite, ${G_{\infty} = 26}$,
    resulting in the contrast between the averaged configurations and
    the extremal configurations.
    However, a large $G_\infty$ suppresses the nucleation rate, 
    ${\Gamma \sim L \exp(- G_\infty)}$,
    and obscures transience in the nucleating droplets.
    To preserve the transient-nucleation regime
    while increasing the Ginzburg parameter, the system size $L$
    must scale exponentially with $G_\infty$.

%
%
%

%
\section{Summary and Discussion}
\label{TheSummary}
    We considered transient nucleation in a long-range
    one-dimensional $\phi^4$ model with dissipative dynamics.
    We defined the metastable well boundary as the set of
    configurations balanced between the metastable and stable phases.
    We estimated the time-dependent likelihood of system
    configurations. We defined the transient critical droplet
    as the most likely configuration constrained to lie on the
    metastable well boundary.  We computed the likelihood of
    the transient critical droplet at various times.
    Our results explain the nonstationarity of
    nucleation rates reported in studies of the time-dependent
    Ginzburg Landau equation.

    As presented here, the theory is applicable to systems near
    metastable equilibrium where the system magnetization evolves slowly
    compared to the relaxation of other dynamical variables,
    as is the case near a spinodal.

    The theory provides a qualitative
    picture of the measured ensemble-averaged profiles without
    any free parameters. The analysis produces a distribution of
    nucleation times consistent with the simulation results with
    a single free parameter.

    Our analysis reduces to the earlier work~\cite{Unger84}
    at long times, when the system has relaxed to
    metastable equilibrium.  We found that
    transient droplets decay to the background
    magnetization in the system at the time of nucleation.
    Before metastable equilibrium, the background magnetization
    acts as an anchor: transient droplets must have greater amplitude, which in turn
    suppresses their rate of formation.
    Furthermore, this suggests that in comparable experiments,
    transient effects result in configurations that are more
    compact than predicted by the metastable
    equilibrium analysis.  Moreover, in systems
    with a stable crystalline phase, these results imply
    transience may determine the
    symmetry of critical droplets when nucleating near a spinodal.

%
%

\medskip

\section{Acknowledgments}
We thank H. Wang whose numerical work prompted
this study, A. Santos for his insightful discussions, and
V. Sood and H. Gould for their careful reading of the manuscript.
We also acknowledge the support of DOE grant DE-FG02-95ER14498.


\begin{thebibliography}{99}
\bibitem{Prefactor}
    The prefactors $\Gamma_0$ and $C$ in
    \myeq{Probability} implicitly depend on time.  For a sufficiently compact
    distribution of nucleation times, both $\Gamma_0$ and $C$ are effectively constant.
\end{thebibliography}
\end{document}